## 2.2. SAGEMATHCLOUD ЯК ЗАСІБ ХМАРНИХ ТЕХНОЛОГІЙ КОМП'ЮТЕРНО-ОРІЄНТОВАНОГО НАВЧАННЯ МАТЕМАТИЧНИХ ТА ІНФОРМАТИЧНИХ ДИСЦИПЛІН

*Метою дослідження* є визначення особливостей здійснення комп'ютерно-орієнтованого навчання математичних та інформатичних дисциплін у хмаро орієнтованому середовищі SageMathCloud. *Задачами дослідження* є дослідження дидактичного потенціалу SageMathCloud, вивчення його структури, виокремлення складових SageMathCloud для створення хмаро орієнтованих програмно-методичних комплексів та дистанційних навчальних курсів з математично-інформатичних дисциплін. *Об'єктом дослідження* є комп'ютерно-орієнтоване навчання математично-інформатичних дисциплін. *Предметом дослідження* є засоби хмарних технологій навчання математично-інформатичних дисциплін. У роботі наведено характеристику дидактичного потенціалу середовища SageMathCloud щодо здійснення комп'ютерно-орієнтованого навчання природничо-математичних та інформатичних дисциплін із використанням засобів хмарних технологій; перелічено та проілюстровано основні компоненти SageMathCloud, що можуть бути використані при проектуванні хмаро орієнтованого програмно-методичного комплексу та дистанційного навчального курсу. *Результати дослідження* будуть покладені в основу написання методичних рекомендацій для викладачів природничих, математичних та інформатичних дисциплін щодо проектування хмаро орієнтованих навчально-методичних комплексів та дистанційних навчальних курсів на базі SageMathCloud.

**Ключові слова**: видавнича система LaTeX, дистанційний навчальний курс, інтерпретатор IPython, система комп'ютерної математики, хмаро орієнтований програмно-методичний комплекс, SageMathCloud.

Одним із найпотужніших засобів хмарних технологій навчання природничо-математичних та інформатичних дисциплін на сьогодні є SageMathCloud – хмаро орієнтований варіант системи комп'ютерної математики SageMath, розміщений на серверах Google [3].

Окрім інтерактивного вивчення зазначених дисциплін, використання SageMathCloud надає такі можливості:

– створення та редагування навчальних і наукових текстів засобами LaTeX, Markdown або HTML;

– співпраця з іншими користувачами в режимі реального часу;

– організація навчальних курсів: додавання студентів, створення власних проектів, моніторингу їх розвитку з використанням хмаро орієнтованих навчально-методичних матеріалів;

– додавання власних файлів, опрацювання даних, оприлюднення результатів та ін.

Основна робота у SageMathCloud відбувається у *проекті* – особистому робочому просторі користувача, в якому зберігаються ресурси різних типів (*.sagews, *.ipynb, *.tex, *.course, *.sage-chat тощо). Кількість незалежних проектів – нерегламентована. У межах спільного проекту користувач-власник має можливість запросити інших до співпраці та оприлюднити окремі файли чи папки.

Кожен проект виконується на сервері SageMathCloud, де він ділить дисковий простір, центральний процесор та оперативну пам'ять з іншими проектами. Безкоштовний тарифний план передбачає використання лише тих ресурсів сервера, що у поточний момент є вільними. Крім того, якщо проект користувача безкоштовного тарифного плану не використовується упродовж декількох тижнів, він переміщується у вторинне сховище з метою вивільнення ресурсів серверу, і його повторний запуск займе суттєво більше часу, ніж у користувача платного тарифного плану.

Учасники проекту можуть об'єднати власні обчислювальні та зберігальні ресурси з метою покращення можливостей проекту в цілому та перерозподілу ресурсів між собою. Організувати спільну роботу з ресурсами проекту SageMathCloud можна або на рівні окремо взятого ресурсу, зокрема робочого аркушу, або на рівні проекту в цілому.

Відкриття спільного доступу на рівні окремо взятого ресурсу є нічим іншим, як web-оприлюдненням вмісту ресурсу у режимі «лише для читання» для всіх користувачів мережі Інтернет, які мають посилання на даний ресурс. Недоліками такого оприлюднення є те, що користувач-«читач» не має можливості управляти обчисленнями на робочому аркуші, навіть якщо автор використав стандартні елементи управління у ньому. Проте, у разі необхідності, оприлюднений робочий аркуш може бути скопійований або завантажений.

Організація спільної роботи на рівні проекту в цілому можлива як без використання ресурсу типу course, так і за його допомогою. Перший спосіб передбачає підключення до проекту учасників, які матимуть можливість спільно працювати з вже існуючими навчальними ресурсами проекту або додавати нові, запрошувати інших учасників, спілкуватись за допомогою текстового та/або відео чатів у рамках спільного проекту. Внесок кожного учасника спільного проекту у вирішення його завдань може бути переглянутий на сторінках історії роботи з проектом або на сторінках його резервних копій [6].

Основними компонентами SageMathCloud для створення *хмаро орієнтованого програмно-методичного комплексу* (рис. 1) з природничо-математичних або інформатичних дисциплін є:

1) *Sage Worksheets* (*.sagews) – робочі аркуші Sage, що надають можливість інтерактивного виконання:

– команд систем комп'ютерної математики (Sage, Axiom, R, Pari, Octave тощо);

– команд мов програмування (C, C#, C++, CoffeeScript, Clojure, Sql, Eiffel, Ecl, Elm, Fortran, Go, Haskell, Java, Julia, JavaScript, Lua, Ocaml, PHP, Perl, Python, Ruby, Scala, Scheme, TypeScript);

– команд мов документування (HTML, Markdown тощо);

2) «блокноти» *IPython* (*.ipynb; з 2016 року – Jupyter Notebook) – синхронізований сеанс мовою програмування Python, який є частиною бібліотеки для наукових та інженерних обчислень SciPy. SageMathCloud надає можливість кільком користувачам взаємодіяти через засоби комунікації у «блокнотах» IPython у синхронному та асинхронному режимах. (Наявність інструменту Besides Sage Worksheets у складі Jupyter Notebooks надає користувачам повний доступ до класичного Linux-терміналу);

3) документи *LaTeX* (*.tex) з повною підтримкою bibtex, sagetex, beamer тощо.

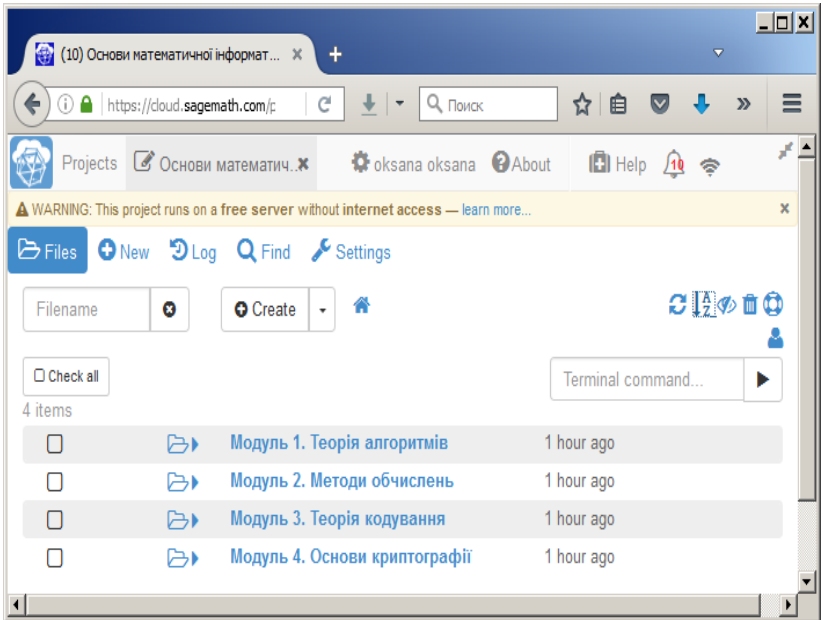

Рис. 1. Сторінка хмаро орієнтованого програмно-методичного комплексу з дисципліни «Математична інформатика» (авт. О. М. Маркова) у SageMathCloud

Оснащеність SageMathCloud *системами реплікації* та *резервного копіювання* є гарантією надійного збереження комплексу-проекту в цілому та окремих його складових, адже збереження кожного проекту здійснюється в трьох фізично відокремлених центрах опрацювання даних, а повне збереження усіх змінених файлів відбувається кожні 2 хвилини [4].

На основі Sage Worksheets можуть бути створені *навчально-методичні матеріали довідникового змісту*, при цьому в одному ресурсі можливе поєднання теоретичних відомостей і інтерактивних прикладів їх застосування на практиці (рис. 2).

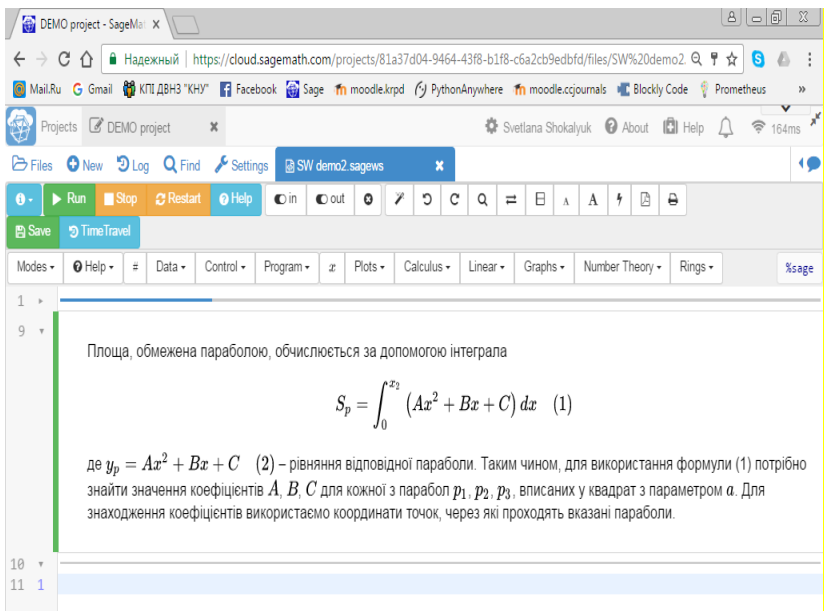

Рис. 2. Фрагмент сторінки інтерактивних навчально-методичних матеріалів довідникового змісту (Sage-аркуш)

За замовчуванням, без додаткових оголошень, на Sage-аркушах можуть бути виконані команди СКМ SageMath. Звернення до команд інших систем комп'ютерної математики можливе після зазначення на початку комірки однієї з так званих «магічних команд» IPython (%axiom, %coffeescript, %cython, %fortran, %gap, %julia, %lisp, %macaulay2, %maxima, %octave, %perl, %python, %r, %scilab, %singular тощо):

```
%r
w <- c(69, 68, 93, 87, 59, 82, 72)
w
plot(w)
```

Окрім навчально-методичних матеріалів довідникового змісту на основі Sage-аркушів можуть бути створені *хмаро орієнтовані інтерактивні робочі зошити* (*практикуми* тощо), призначені для формування й розвитку умінь застосовувати набуті теоретичні знання на практиці (рис. 3).

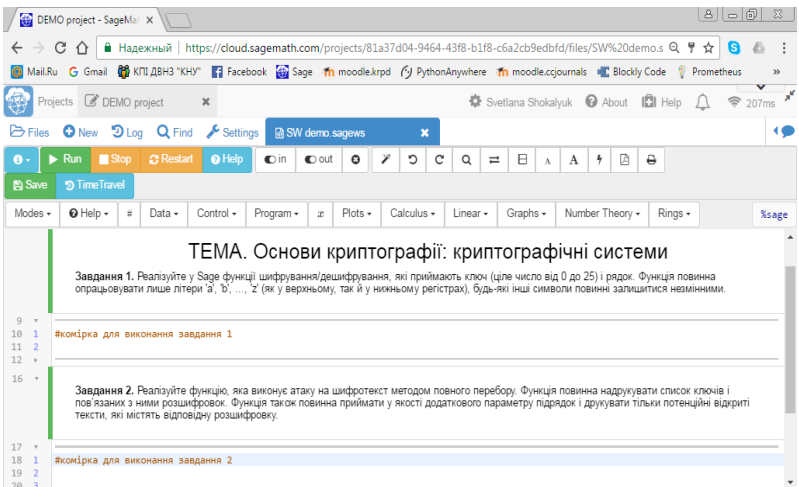

Рис. 3. Фрагмент сторінка хмаро орієнтованого робочого зошиту (Sage-аркуш)

Щодо хмаро орієнтованої підтримки практикумів з інформатичних дисциплін, зокрема програмування, більш зручним і неперевантаженим додатковим математичним інструментарієм буде ресурс *Jupyter блокнот* (рис. 4).

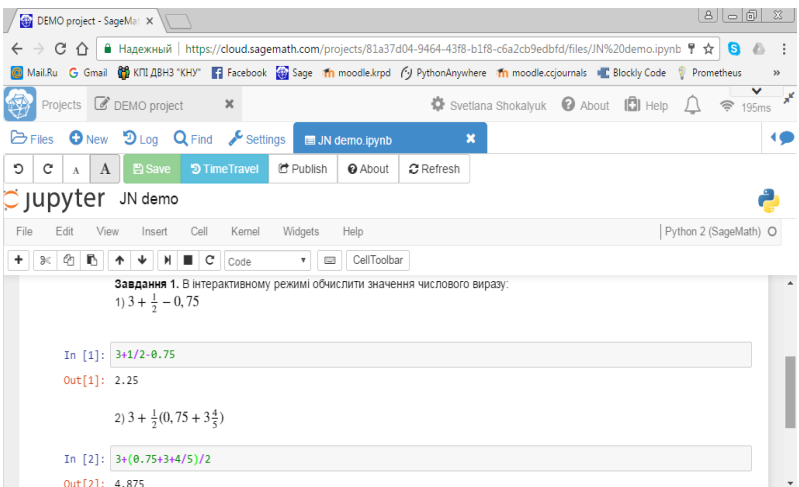

Рис. 4. Фрагмент сторінки хмаро орієнтованого робочого зошиту (Jupyter блокнот)

Підготовка якісних навчально-математичних або науково-математичних текстів (як статичного, так і динамічного змісту) до друку (або зручної роботи із документацією у pdf-форматі) може бути виконана за допомогою інструментарію системи *LaTeX*, як складової SageMathCloud.

LaTeX – потужна видавнича система, що відома надзвичайною стабільністю, здатністю працювати на багатьох комп'ютерних платформах і операційних системах, а також практично повною відсутністю помилок. Номер версії LaTeX збігається до числа π і зараз дорівнює 3.14159 [2].

Для розробки україномовних математичних текстів у LaTeX необхідно підключити модулі для підтримки українського правопису та кодування Unicode. Наприклад (рис. 5):

```
%описова частина документа - преамбула
\documentclass[a5paper,10pt]{article}
\usepackage[ukrainian]{babel}
\usepackage[utf8]{inputenc}
\usepackage{sagetex}
\title{Навчально-методичний посібник
\\"SageMathCloud у навчанні природничо-математичних
та інформатичних дисциплін"}
\author{С.В. Шокалюк, О.М. Маркова,
С.О. Семеріков\\
Спільна лабораторія з питань використання хмарних
технологій в освіті}
\date{2017}
%тіло документу
\begin{document}
\maketitle
\end{document}
```

Можливість підготовки математичних текстів динамічного змісту (із результатами обчислень) надають вбудовані у tex-документ команди спеціалізованого пакету для системи LaTeX – SageTeX [5].

Для початку використання SageTeX слід вказати \usepackage{sagetex} в описовій частині tex-документа. Для вставки результатів виконання Sage-команд у tex-документ застосовується команда \sage{<код Sage>}, де <код

`Sage>` – будь-який код мовою Sage. Наприклад, виконання команди `\sage{matrix([[1,2], [3,4]])^2}` призведе до появи у тексті її результату мовою LaTeX:

```
\left(\begin{array}{rr}
7 & 10 \\
15 & 22
\end{array}\right)
```

У команді `\sage` також можуть бути використані посилання на змінні Sage, визначені раніше у поточному документі.

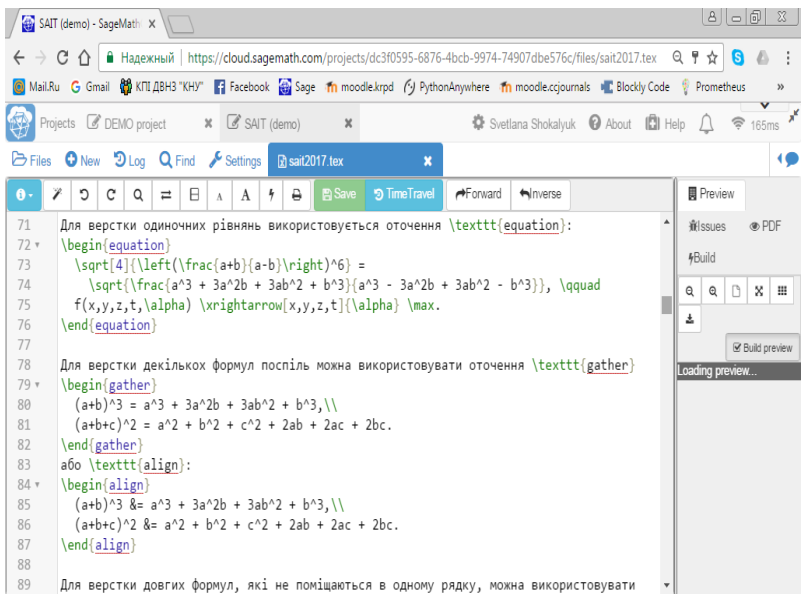

Рис. 5. Фрагмент сторінки із макетом друкованого посібника математичного змісту (tex-документ)

Графічні побудови виконуються за допомогою команди `\sageplot[<опції LaTeX>][<формат>]{<графічні об'єкти>, <ключові аргументи>}`, де:

<опції LaTeX> – будь-який текст, що передається у необов'язкові аргументи, вказані в квадратних дужках, команди `\includegraphics`. Якщо не вказано, буде

137

використовуватися \width=.75\textwidth;

<формат> – вказує розширення графічного файлу, у якому Sage намагатиметься зберегти рисунок. Якщо не вказано, буде збережено у форматі EPS або PDF;

<графічні об'єкти> – відповідні графічні команди мовою Sage;

<ключові аргументи> – будь-які параметри виконання попередніх.

Наприклад, команда \sageplot[angle=30, width=5cm] {plot(sin(x),0,pi),axes=False} призведе до появи у тексті її результату – зображення частини синусоїди, нахиленої під кутом 30°.

Якщо у команді була помилка, замість результату її виконання ageTeX вставляє два знаки питання.

Для тривимірної графіки зазначення формату вихідного графічного файлу є обов'язковим:

```
\sage{var("x y")}
\sageplot[][png]{plot3d(sin(x+y), [x, 0, pi],\
[y, -pi,0]), axes=False}
```

Останній приклад можна було б оформити у вигляді фрагменту мовою Sage як блок за допомогою команд \begin{sageblock} та \end{sageblock}. Наприклад, можна визначити блок:

```
\begin{sageblock}
   var('x')
   f(x) = sin(x) - 1
   g(x) = log(x)
   h(x) = diff(f(x) * g(x), x)
\end{sageblock}
```

Після цього у тексті можна звертатись до описаних у ньому функцій та змінних. Наприклад, «Маємо $h(2) = \sage{h(2)}$, де $h$ є похідною добутку $f$ та $g$». Виклик \sage буде коректно замінений на sin(1)-1.

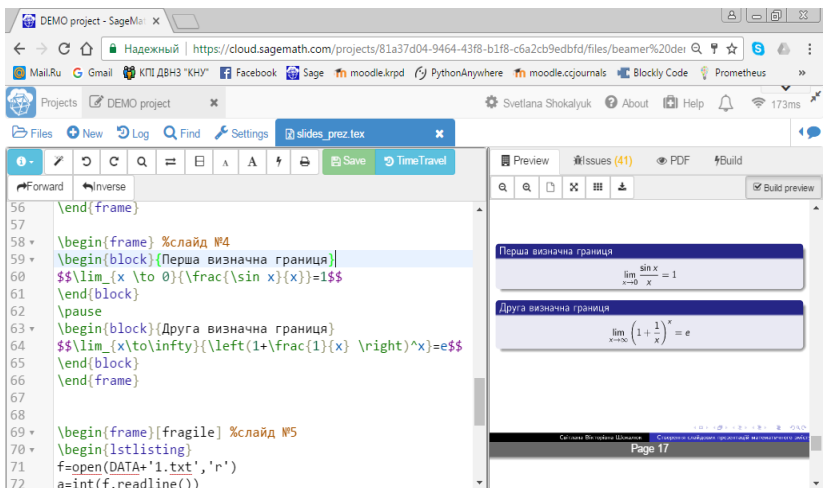

Рис. 6. Фрагмент сторінки із макетом слайдової презентації математичного змісту (tex-документ)

У LaTeX-редакторі на основі шаблону beamer – \documentclass{beamer} – можуть бути створені якісні *слайдові презентації математичного змісту* (рис. 6).

Налаштування дизайну презентації (стиль та схема) визначається за допомогою команд \usetheme{назва_стилю_оформлення} та \useoutertheme{назва_кольорової_схеми_презентації}.

Слайди презентації додаються у документ за допомогою команд-оточення frame:

```
\begin{document}
%додавання титульного слайду презентації
\begin{frame}
   \titlepage
\end{frame}
%додавання нового слайду
\begin{frame}
   %вміст змістового слайду №1
\end{frame}
%додавання нового слайду
\begin{frame}
   %вміст змістового слайду №2
```

```
\end{frame}
\end{document}
```

Слайд без «шапки» й «підвалу» створюється із зазначенням додаткового параметра plain:

```
\begin{frame}[plain]
   %вміст слайду без «шапки» й «підвалу»
\end{frame}
```

Для додавання слайду, що містить лістинг програмного коду необхідно додатково підключити пакет listings (\usepackage{listings} в преамбулі документа), а сам фрагмент програмного коду на слайді розмістити в тілі команди lstlisting:

```
\begin{frame}[fragile]
\begin{lstlisting}
 f=open(DATA+'1.txt','r')
 a=int(f.readline())
 b=int(f.readline())
 print "(%d,%d)"%(a,b)
 while a<>b:
    if a>b:
        a=a-b
    else:
        b=b-a
    print "(%d,%d)"%(a,b)
 f.close()
\end{lstlisting}
\end{frame}
```

Для виділення важливої частини тексту на слайді призначені команди-оточення block, alertblock та exampleblock:

```
\begin{block}{заголовок синього блоку}
   %вміст синього блоку
\end{block}
\begin{alertblock}{заголовок червоного блоку}
   %вміст червоного блоку
\end{alertblock}
\begin{exampleblock}{заголовок зеленого блоку}
   %вміст зеленого блоку
```

\end{exampleblock}

Ілюзія анімації об'єктів на слайді та переходів між слайдами може бути створена за допомогою команди \pause:

```
\begin{frame} %слайд №4
\begin{block}{Перша визначна границя}
$$\lim_{x \to 0}{\frac{\sin x}{x}}=1$$
\end{block}
\pause
\begin{block}{Друга визначна границя}
$$\lim_{x\to\infty}{\left(1+\frac{1}{x}
\right)^x}=e$$
\end{block}
\end{frame}
```

Приклад LaTeX-коду з проектуванням анімації елементів списку на слайді:

```
\begin{frame} %слайд
\begin{block}{Стилі презентації}
\begin{itemize}
\item<1-> CambridgeUS
\item<2-> Hannover
\item<3-> Madrid
\item<4-> Warsaw
\item<5-> інші
\end{itemize}
\end{block}
\end{frame}
```

Додавання інших об'єктів на слайд – таблиць, формул, графічних зображень тощо – здійснюється звичним чином.

Отже, хмаро орієнтований програмно-методичний комплекс, створений у SageMathCloud як структурований набір Sage-аркушів, Jupyter блокнотів та/або tex-документів (макетів друкованих та презентаційних навчально-методичних матеріалів математичного змісту) є основою комп'ютерно-орієнтованого навчання природничо-математичних та інформатичних дисциплін у форматі електронного навчального курсу, в якому управління та контроль навчанням здійснюється опосередковано. Повноцінне дистанційне навчання у

SageMathCloud може бути організоване за допомогою *менеджера курсів* (Manage a Course), в якому структуровані (розподілені за папками) складові хмаро орієнтованого програмно-методичного комплексу є його визначальними змістовими елементами [6].

**Shokaljuk S. V., Markova O. M., Semerikov S. O. SageMathCloud as the Learning Tool Cloud Technologies of the Computer-Based Studying Mathematics and Informatics Disciplines**

The *aim* of this study is to determine peculiarities of computer-based learning and mathematical disciplines in informatychnyh oriented cloud environment SageMathCloud. *The objectives of the study* is to investigate the didactic potential SageMathCloud, study its structure SageMathCloud isolation components to create a cloud-oriented software and learning systems and distance learning courses in the mathematics and informatics disciplines. The *object of research* is computed mathematically-oriented education informatics disciplines. The *subject of research* is the learning tool cloud technologies of the computer-based studying mathematics and informatics disciplines. In this paper, the characteristics of the didactic potential SageMathCloud environment for the implementation of computer-based learning natures, mathematics and informatics disciplines with the use of cloud technologies; lists and illustrates the basic components SageMathCloud, which can be used in the design of cloud-based software and methodical complex and remote training course. *Results of the study* will be the basis for writing guidelines for teachers of natural, mathematics and informatics cloud-oriented software and learning systems and distance learning courses based on SageMathCloud.

**Keywords**: cloud-oriented program-methodical complex, distance learning course, IPython interpreter, mathematics computer system, publishing LaTeX, SageMathCloud.